\documentclass[prl,twocolumn,showpacs]{revtex4}
\usepackage{graphicx}
\usepackage{amsmath,amssymb}
\usepackage[colorlinks,citecolor=blue]{hyperref}

\begin{document}

\title{Classification of quantum critical states of integrable
antiferromagnetic spin chains and their correspondent two-dimensional
topological phases}
\author{Zheng-Xin Liu$^{1}$ and Guang-Ming Zhang$^{2}$}
\email{gmzhang@tsinghua.edu.cn}
\affiliation{$^{1}$Institute for Advanced Study, Tsinghua University, Beijing 100084,
China.\\
$^{2}$State Key Laboratory of Low-Dimensional Quantum Physics, Department of
Physics, Tsinghua University, Beijing 100084, China}
\date{\today}

\begin{abstract}
We examine the effective field theory of the Bethe ansatz integrable
Heisenberg antiferromagnetic spin chains. It shows that the quantum critical
theories for the integer spin-$S$ chains should be characterized by the $%
SO(3)$ level-$S$ Wess-Zumino-Witten model, and classified by the third
cohomology group $H^{3}(SO(3),Z)=Z$. Depending on the parity of spin $S$,
this integer classification is further divided into two distinct
universality classes, which are associated with two completely different
conformal field theories: the even-$S$ chains have gapless bosonic
excitations and the odd-$S$ chains have both bosonic and fermionic
excitations. We further show that these two classes of critical states
correspond to the boundary states of two distinct topological phases in two
dimension, which can be described by two-dimensional doubled $SO(3)$
topological Chern-Simons theory and topological spin theory, respectively.
\end{abstract}

\pacs{05.30.-d, 05.70.Jk, 75.10.Kt}
\maketitle

\textit{Introduction}.- The study of topological phases and their
classification has become an important issue in condensed matter physics.
Historically, topological order\cite{Wen-Niu} was proposed to describe
fractional quantum Hall states,\cite{Tsui-1982,Laughlin-1983} which cannot
be characterized by a local order parameter with spontaneous symmetry
breaking. The nature of these states is long-range entanglement and a finite
degeneracy of the ground states on a torus.\cite{Kitaev-1997,chen-gu-wen}
Recently, it is shown that some nontrivial properties, such as robust
gapless edge excitations, exist even in quantum gapped states with only
short-ranged entanglement, known as symmetry protected topological (SPT)
phases. The best example is the state of topological insulators\cite%
{Hasan-Kane,Qi-Zhang} protected by time reversal symmetry and charge
conservation symmetry. Actually, such SPT phases also exist in
one-dimensional quantum antiferromagnetic spin chains.\cite{chen-gu-wen}

More than thirty years ago, from the theory of nonlinear sigma model with $%
\theta $ term, Haldane\cite{Haldane-1983} predicted that antiferromagnetic
Heisenberg spin chains are classified into two universality classes:
half-odd integer spins with gapless excitations and integer spins with
gapped excitations. Recent studies\cite{Gu-Wen-2009,Pollmann-2012} indicated
that the Haldane gapped phase for an odd integer spin chain is a nontrivial
SPT phase, while the even integer spin chain is trivial, because the edge
states are not symmetry protected. The number of SO(3) symmetry protected
distinct topological phases can be labeled by the elements of the second
group cohomology of $SO(3)$ group\cite{Chen-Gu-Wen1D,chen-gu-liu-wen}: $%
\mathcal{H}^{2}(SO(3),U(1))=Z_{2}$. In order to find a complete
classification scheme including the critical states of the half-odd-integer
spin antiferromagnetic chains, it is desirable to consider the problem as
whether the differences between the odd and even integer Haldane gapped
phases can be understood from their effective field theories. Moreover the
classification of one-dimensional quantum critical states is of itself an
important issue, because of its close relation to the boundary theories of
two-dimensional topological ordered phases.

In this Letter, we will consider this issue using the Bethe ansatz
integrable antiferromagnetic spin chains with arbitrary spin.\cite%
{Takhtajan,Babudjian} It has been known that the critical states for the
half-odd integer spins are described by the $SU(2)$ level-$2S$
Wess-Zumino-Witten (WZW) model.\cite%
{Affleck-1986,Affleck-Haldane,Affleck-1989,Avdeev-1990} Realizing that it is
the $SO(3)$ group symmetry as the faithful representation for integer spins,
we point out that the quantum critical states for the integer spin chains
should be regarded as the $SO(3)$ level-$S$ WZW models, which correspond to
a two-dimensional (doubled) $SO(3)$ topological spin gauge theories\cite%
{Dijkgraaf-Witten}. Depending on the parity of spins, there exist two
universality classes with completely different conformal field theories
(CFTs): the even integer chains have bosonic excitations and the odd integer
chains have both bosonic and fermionic excitations. Such a classification is
consistent with the classification for the 1D $SO(3)$ gapped phases from the
second group cohomology,\cite{Chen-Gu-Wen1D,chen-gu-liu-wen} and is further
shown to have an intimate relation to 2D $SO(3)$ SPT phases or chiral spin
liquid phases.

\textit{Model Hamiltonian}.- We begin with the $SU(2)$ quantum spin chains,
where the spin operators on each site take values in the spin $su(2)$
algebra. The relevant irreducible representations are labeled by the spin $%
S\in \{1/2,1,3/2,...\}$, including the half-odd integer and integer spins.
The standard one-dimensional spin-$S$ antiferromagnetic Heisenberg model is
given by%
\begin{equation}
H=J\sum_{n=1}^{L}\mathbf{S}_{n}\cdot \mathbf{S}_{n+1},\text{ }%
S_{n}^{2}=S(S+1),\text{ \ }J>0.  \label{HSM}
\end{equation}%
For the $S=1/2$ case, a quantum critical phase was obtained by the Bethe
ansatz solution, and spin-spin correlation functions show power law
behavior. However, no exact solution exists for $S>1/2$.

For the $S=1$ case, a general $SU(2)$ symmetric model has the form
\begin{equation}
H=\sum_{n=1}^{L}\left[ \left( \cos \theta \right) \mathbf{S}_{n}\cdot
\mathbf{S}_{n+1}+\left( \sin \theta \right) \left( \mathbf{S}_{n}\cdot
\mathbf{S}_{n+1}\right) ^{2}\right] ,  \label{BLBQ}
\end{equation}%
where the Haldane gapped phase has been confirmed in the parameter range $%
-\pi /4<\theta <\pi /4$, and the dimerized phase is for $-3\pi /4<\theta
<-\pi /4$. There is a quantum critical point at $\theta =-\pi /4$, which is
also exactly solved by the Bethe ansatz method.\cite{Takhtajan,Babudjian} We
can understand the nature of both the Haldane gapped phase and dimerized
phase from the critical field theory, which has been characterized by the $%
SU(2)$ level-$2$ or $SO(3)$ level-$1$ WZW model in term of three free
Majorana fermions. Depending on the sign of the relevant mass term, an
energy gap is generated for the Haldane phase and dimerized phase,
respectively.\cite{Tsvelik-1990}

Actually, there is a generalization of the antiferromagnetic Heisenberg spin
chain to arbitrary spin-$S$ with preserving the $SU(2)$ symmetry and
integrability. The model Hamiltonian can be defined by\cite%
{Takhtajan,Babudjian}%
\begin{equation}
H_{S}=J\sum_{n=1}^{L}Q_{2S}(\mathbf{S}_{n}\cdot \mathbf{S}_{n+1}),
\end{equation}%
where $S_{n}=(S_{n}^{x},S_{n}^{y},S_{n}^{z})$ is the $SU(2)$ generator of
arbitrary integer or half-odd integer spin, and the special polynomial of
degree $2S$ is%
\begin{equation*}
Q_{2S}(X)=-\sum_{j=1}^{2S}\left( \sum_{k=1}^{j}\frac{1}{k}\right)
\prod_{l=0,l\neq j}^{2S}\left( \frac{X-X_{l}}{X_{j}-X_{l}}\right)
\end{equation*}%
with $X_{j}=\frac{1}{2}\left[ j(j+1)-2S(S+1)\right] $. For $S=1/2$, the
Hamiltonian $H_{1/2}$ is the antiferromagnetic Heisenberg spin model (\ref%
{HSM}), while the Hamiltonian for $S=1$ corresponds to the model (\ref{BLBQ}%
) at the quantum critical point $\theta =-\pi /4$.

\textit{Critical field theories.}- Generally, one can expect that the Bethe
ansatz integrable Hamiltonians with arbitrary spin describe a family of
quantum critical states with the $SU(2)$ symmetry, which can be described by
(1+1)D CFTs. When the conformal central charge $c=3S/(S+1)$ and
thermodynamic properties are compared with the Bethe ansatz results for
several small spin cases,\cite{Affleck-1986,Affleck-1989,Avdeev-1990} the
effective critical field theory had been suggested by the $SU(2)$ level-$2S$
WZW model with the action:%
\begin{eqnarray}
S(g) &=&-\frac{ik}{12\pi }\int d^{3}y\epsilon ^{\mu \nu \lambda }\text{Tr}%
\left[ g^{-1}\partial _{\mu }gg^{-1}\partial _{\nu }gg^{-1}\partial
_{\lambda }g\right]  \notag \\
&&+\frac{k}{8\pi }\int dxd\tau \text{Tr}\left( \partial _{\mu }g\partial
_{\mu }g^{-1}\right) ,  \label{SU2}
\end{eqnarray}%
where $g\in SU(2)$ is the group element and $k=2S$ is the level index. When
a closed boundary condition is imposed, the first term is the (1+1)D WZW
action. Actually, the above (1+1)D WZW model can be considered as the
boundary theory of the following (2+1)D principal chiral model\cite%
{Xu-Ludwig}
\begin{eqnarray}
S_{\mathrm{PC}} &=&\int d^{2}xd\tau \left[ {\frac{1}{4\kappa ^{2}}}\mathrm{Tr%
}(\partial _{\mu }g\partial _{\mu }g^{-1})\right.  \notag \\
&&\left. -{\frac{ik}{12\pi }}\varepsilon ^{\mu \nu \lambda }\mathrm{Tr}%
(g^{-1}\partial _{\mu }gg^{-1}\partial _{\nu }gg^{-1}\partial _{\lambda }g)%
\right]  \label{PCSU2}
\end{eqnarray}%
with the coupling constant $\kappa ^{2}\rightarrow \infty $ in the strong
coupling limit, where the second term is quantized on the closed space-time
manifold. When the space is open, the second term becomes an effective WZW
term for the boundary theory and flows to the fixed point (\ref{SU2}).

However, there exist subtleties related to the symmetry group, which has
been neglected in the previous critical effective field analysis.\cite%
{Affleck-1986,Affleck-Haldane,Affleck-1989,Avdeev-1990} When we lift the
symmetry described in terms of the $su(2)$ Lie algebra to a group symmetry
of $SU(2)$, we might have several choices, and not all of them will lead to
an $SU(2)$ faithful representation.\cite{chen-gu-liu-wen} If the physical
spins transform in half-odd integer spin representations, the group $SU(2)$
is acting faithfully. However, if the physical spins transform in integer
spin representations, the $SU(2)$ does not act faithfully and the actual
symmetry should be $SO(3)=SU(2)/Z_{2}$. The $SU(2)$ is a two-fold covering
of the $SO(3)$. Moreover, it is important to understand the differences
between $SU(2)$ and $SO(3)$ more precisely. When viewed as geometric
manifolds, $SU(2)$ and $SO(3)$ look \textit{locally} identical. However,
they differ in their \textit{global} topology: $SU(2)$ is simply-connected
and the first homotopy group $\pi _{1}(SU(2))=0$; while $SO(3)$ is
non-simply connected, $\pi _{1}(SO(3))=Z_{2}$, namely, it admits non-trivial
loops which cannot be contracted to a point.

Therefore, for the integrable half-odd integer spin chains, it is correct to
identify the effective field theory as the $SU(2)$ level-$2S$ WZW model.
However, for the quantum critical states of the integer spin chains,
although the effective action has the same conformal central charge, the
corresponding effective field theory should be regarded as the $SO(3)$ level-%
$S$ WZW model. The corresponding WZW action is given by%
\begin{eqnarray}
S^{\prime }(g) &=&-\frac{iS}{24\pi }\int d^{3}y\epsilon ^{\mu \nu \lambda }%
\text{Tr}\left[ g^{-1}\partial _{\mu }gg^{-1}\partial _{\nu }gg^{-1}\partial
_{\lambda }g\right]  \notag \\
&&+\frac{S}{16\pi }\int dxd\tau \text{Tr}\left( \partial _{\mu }g\partial
_{\mu }g^{-1}\right) ,  \label{SO3}
\end{eqnarray}%
where $g\in SO(3)$ as the group elements. Similarly, the above (1+1)D WZW
model can also be considered as the boundary theory of the following (2+1)D $%
SO(3)$ principal chiral model
\begin{eqnarray}
S_{\mathrm{PC}}^{\prime } &=&\int d^{2}\mathbf{x}d\tau \left[ {\frac{1}{%
8\kappa ^{2}}}\mathrm{Tr}(\partial _{\mu }g\partial _{\mu }g^{-1})\right.
\notag \\
&&\left. -{\frac{iS}{24\pi }}\varepsilon ^{\mu \nu \lambda }\mathrm{Tr}%
(g^{-1}\partial _{\mu }gg^{-1}\partial _{\nu }gg^{-1}\partial _{\lambda }g)%
\right] ,  \label{PCSO3}
\end{eqnarray}%
which has a global $SO(3)_{L}\times SO(3)_{R}$ symmetry.\cite{note1} Namely,
the action $S_{\mathrm{PC}}^{\prime }$ is invariant under the transformation
$g(\tau ,\mathbf{x})=\Omega g(\tau ,\mathbf{x})\bar{\Omega}^{-1}$, where $%
\Omega \in SO(3)_{L}$ and $\bar{\Omega}\in SO(3)_{R}$ are independent
constant group elements. However, in the boundary theory of (\ref{SO3}), an
enhanced local symmetry exists $g(\tau ,x)\rightarrow \Omega (\tau +x)g(\tau
,x)\bar{\Omega}^{-1}(\tau -x)$.

\textit{Excitation spectra from CFT}.- So far the obtained results for
integer spin $S$ seem to indicate that the $SO(3)$ level-$S$ WZW model are
refinements of the $SU(2)$ level-$2S$ WZW model. Actually, the critical
field theories with odd and even integer spins represent two distinct
universality classes of the quantum critical states, which are completely
different from those quantum critical theories of the $SU(2)$ level-$2S$ WZW
model. By performing modular invariance to the $SO(3)$ level-$S$ WZW model%
\cite{Gepner-Witten-1986} or constructing the space of states in the
canonical quantized theory\cite{Felder-1988}, the CFT spectra have been
worked out in terms of the primary fields ($j,\overline{j}$), where $j$ and $%
\overline{j}$ are the left $SO(3)_{L}$ and right $SO(3)_{R}$ isospin quantum
number.
\begin{table}[t]
\caption{Conformal field theory spectra of the $SU(2)_{2S}$ and $SO(3)_{S}$
WZW models. The central charge of both models is given by $c=3S/(S+1)$.}
\label{tab:spectrum}%
\begin{ruledtabular}
\begin{tabular}{ccc}
WZW model & & Primary fields ($j,\bar j$) \ \ \ \ \ \ \ \ \ \ \ \  \\
\hline
$SU(2)_{2S}$ & & (0,0), ($1\over2$,$1\over2$),(1,1),...,($S,S$)\ \ \ \ \ \ \ \ \ \ \ \ \  \\
\hline
$SO(3)_{S}$ & untwisted: & (0,0), ($1,1$),...,($S,S$)\ \ \ \ \ \ \ \ \ \ \ \ \ \ \ \ \ \ \ \ \ \\
($S$=even)  & twisted: \ \ \ \ & (0,S), ($1,S-1$),...,($S,0$)\ \ \ \ \ \ \ \ \ \ \ \ \ \ \ \ \\
\hline
$SO(3)_{S}$& untwisted: & (0,0), ($1,1$),...,($S,S$)\ \ \ \ \ \ \ \ \ \ \ \ \ \ \ \ \ \ \ \ \ \\
($S$=odd)   & twisted: \ \ \ \ & (${1\over2},S-{1\over2}$), (${3\over2},S-{3\over2}$),...,($S-{1\over2},{1\over2}$)\\
\end{tabular}
\end{ruledtabular}
\end{table}

For the $SO(3)$ level-$S$ WZW model with \textit{even} integer $S$, the
corresponding CFT includes the primary fields, consisting of the untwisted
and twisted sectors with multiplicity one.\cite%
{Gepner-Witten-1986,Felder-1988} (see Tab.\ref{tab:spectrum}) Since all
these primary fields have integer left and right isospin quantum numbers,
they represents \textit{bosonic} fields with integer conformal weights,
forming a chiral algebra. The field $(0,0)$ denotes the identity $I$ and
some primary fields appearing in both sectors correspond to two different
primary fields which have the same transformation properties under the
current algebra. The field $(1,1)$ in the untwisted sector is nothing but
the $SO(3)$ symmetric elementary field $g(\tau ,x)$, and the operator
products of $g$ with itself generate all the primary fields in the untwisted
sector. The primary fields in the twisted sector are so-called spin fields,
which are generated by the elementary soliton field $(0,S)$. There is an
important selection (fusion) rule: $(0,S)\times (0,S)=I$. Actually, the
operator products of the elementary and soliton fields with itself or
sufficient number of times can give rise to all the primary fields. So they
represent elementary fields of the critical field theory.\cite%
{Gepner-Witten-1986,Felder-1988}

For the $SO(3)$ level-$S$ WZW model with \textit{odd} integer $S$, the
primary fields also consist of the untwisted sector and twisted sector with
multiplicity one.\cite{Gepner-Witten-1986,Felder-1988} (see Tab.\ref%
{tab:spectrum}) Since the twisted primary fields contain half-odd integer
isospin quantum numbers, the conformal weights of these primary fields
include both half-odd-integers (\textit{fermionic}) and integers (\textit{%
bosonic}), forming a $Z_{2}$ graded chiral algebra or chiral super-algebra.%
\cite{Dijkgraaf-Witten} Moreover, the selection (fusion) rule for the
product of the elementary spin field with itself becomes\cite%
{Gepner-Witten-1986}%
\begin{equation}
(\frac{1}{2},S-\frac{1}{2})\times (\frac{1}{2},S-\frac{1}{2})=I+(1,1),
\end{equation}%
which indicates that the elementary primary field may be generated from the
operator product of the elementary spin field with itself.

Therefore, the $SO(3)$ level-$S$ WZW model is divided into two universality
classes: one is denoted by the even level index with chiral algebra and
another corresponds to the odd level index with chiral super-algebra. Such a
classification is consistent with the analysis of the D-brane charge groups,%
\cite{GabeGannon} where the even level theories have a charge group $%
Z_{2}\times Z_{2}$, while the odd level theories have a charge group $Z_{4}$.

As a comparison, the CFT spectrum of the $SU(2)$ level-$2S$ WZW model is
also given in Tab.\ref{tab:spectrum}, where a single sector includes the
primary fields with both integer and half-odd integer isospin quantum
numbers with multiplicity one.\cite{Felder-1988} However, these primary
fields can only form the chiral algebra with integer conformal weights.
Interestingly, there is a coincidence for the critical state of the $S=1$
chain. The obtained CFT primary fields $(0,0)$, $(\frac{1}{2},\frac{1}{2})$,
and $(1,1)$ can be given by both the $SO(3)$ level-$1$ WZW model and the $%
SU(2)$ level-$2$ WZW model.

Away from the critical point of the integer spin chains, the elementary spin
field is not allowed by the symmetry, and the most relevant operator is
given by the elementary field of the WZW model: $\mathcal{L}_{\text{int}%
}=-\lambda $Tr$g$, which corresponds to a mass term of the critical field
theory. Such a mass term can produce an energy gap in the spin excitations,
leading to the Haldane gapped phase ($\lambda >0$) or the dimerized phase ($%
\lambda <0$), respectively. However, for the $SU(2)$ level-$2S$ WZW model,
the above mass operator does not permit, so the leading relevant operator
should be given by $\mathcal{L}_{\text{int}}=-\lambda \left( \text{Tr}%
g\right) ^{2}$ in terms of $g\in SU(2)$.

\textit{Correspondent 2D topological phases}.- According to the work of
Dijkgraaf and Witten,\cite{Dijkgraaf-Witten} the (2+1)D topological
Chern-Simons theories with gauge group $G$ are classified by the fourth
cohomology group $H^{4}(BG,Z)$, where $BG$ is the classifying space of the
group $G$. For a non-simply connected gauge group $G$, namely, $\pi
_{1}(G)\neq 0$, the level-$k$ can not take arbitrary integer values.
However, for the non-simply connected group $G=SO(3)$, $k$ should be an even
integer.\cite{MooreSeiberg,Dijkgraaf-Witten} Notice that our definition of
the level $k$ for $SO(3)$ is half of that defined in Ref.\cite%
{MooreSeiberg,Dijkgraaf-Witten}. Compare to these references, we should
redefine $k\rightarrow 2k$. More importantly, the (2+1)D $SO(3)$
Chern-Simons topological theory has a correspondence to a (1+1)D $SO(3)$ WZW
model with even level $k$, and there exists an inverse transgression map: $%
H^{4}(BG,Z)\rightarrow H^{3}(G,Z)$.

However, since the (2+1)D space-time manifold always support a spin
structure, the level $k$ can take more integer values. For $G=$ $SO(3)$, the
level-$k$ can be either even or odd integers. The corresponding $SO(3)$
topological gauge theories with an odd level-$k$ are called topological spin
theories, and also have the correspondence to the (1+1)D $SO(3)$ WZW model
with the same level-$k$, which is however associated to a Z$_{2}$ graded
chiral algebra or chiral super-algebra.\cite{Dijkgraaf-Witten}

Next we will show how the principal chiral model (\ref{PCSO3}) can provide a
bridge to connect the (2+1)D $SO(3)$ topological spin Chern-Simons theory to
the (1+1)D $SO(3)$ WZW model with the same level index. To this end, we use
the method of ''gauging'' the symmetry group. Because the model (\ref{PCSO3}%
) has $SO(3)_{L}\times SO(3)_{R}$ symmetry, we can regard this symmetry
group as a gauge group and minimally couple to two external gauge fields $%
A_{\mu }$ and $\bar{A}_{\mu }$. Namely we replace $g^{-1}\partial _{\mu }g$
by $g^{-1}(\partial _{\mu }+A_{\mu })g$ and $g\partial _{\mu }g^{-1}$ by $%
g(\partial _{\mu }+\bar{A}_{\mu })g^{-1}$, respectively.\cite{note2} After
integrating out the group elements, an effective action can be derived
\begin{eqnarray}
S(A,\overline{A}) &=&{\frac{ik}{8\pi }}\int_{M}d^{3}x\varepsilon ^{\mu \nu
\lambda }\left[ (A_{\mu }^{a}\partial _{\nu }A_{\lambda }^{a}+{\frac{%
i\varepsilon _{abc}}{3}}A_{\mu }^{a}A_{\nu }^{b}A_{\lambda }^{c})\right.
\notag \\
&&\left. -(\bar{A}_{\mu }^{a}\partial _{\nu }\bar{A}_{\lambda }^{a}+{\frac{%
i\varepsilon _{abc}}{3}}\bar{A}_{\mu }^{a}\bar{A}_{\nu }^{b}\bar{A}_{\lambda
}^{c})\right] ,  \label{DCS}
\end{eqnarray}%
which is a doubled Chern-Simons topological gauge theory on a $SO(3)$ spin
manifold $M$.\cite{Freedman-2003} Since the two external gauge fields $%
A_{\mu }$ and $\bar{A}_{\mu }$ act independently, it is natural to expect no
cross terms appearing in this action.

To justify the self-consistency of our results, we reverse the reasoning of
the previous discussion, and derive the (1+1)D WZW action from the doubled
Chern-Simons action (\ref{DCS}). First of all, the action (\ref{DCS}) is
invariant under time reversal transformation: $i\rightarrow -i$, $%
t\rightarrow -t$, $\tau \rightarrow \tau $, and $A_{\mu }\rightarrow \bar{A}%
_{\mu }$. Following the method of Ref.\cite{MooreSeiberg}, we can choose a
specific gauge and integrate out $A_{0}$, resulting in a delta function $%
\delta (F_{ij})$ with $F_{ij}=\partial _{i}A_{j}-\partial _{j}A_{i}+{\frac{1%
}{2}}[A_{i},A_{j}]$. Then substituting the solution $A_{i}=g\partial
_{i}g^{-1}$ into the first part of (\ref{DCS}), we obtain a \textit{chiral}
version of a (1+1)D $SO(3)$ level-$k$ WZW model,
\begin{eqnarray}
S_{1} &=&{\frac{ik}{48\pi }}\int_{M}d^{3}x\epsilon ^{\mu \nu \lambda }%
\mathrm{Tr}(g^{-1}\partial _{\mu }gg^{-1}\partial _{\nu }gg^{-1}\partial
_{\lambda }g)  \notag \\
&&+{\frac{ik}{16\pi }}\int_{\partial M}dx_{1}dx_{0}\mathrm{Tr}(\partial
_{1}g^{-1}\partial _{0}g),
\end{eqnarray}%
where $x_{1}$ is the spacial coordinate of the boundary and $x_{0}=\tau $.
Noticing that $\bar{A}_{i}$ is the time reversal partner of $A_{i}$, we can
set $\bar{A}_{i}=g^{-1}\partial _{i}g$, where we have assumed that this
Lagrangian is strictly invariant under time reversal, and then another
action $S_{2}$ with \textit{opposite} chirality can be derived from the
second part of (\ref{DCS}). Compared to $S_{1}$, the second term in $S_{2}$
has a negative sign only, and the first term in $S_{2}$ is exactly the same.
Finally the total action is thus deduced
\begin{equation}
S_{\mathrm{top}}={\frac{ik}{24\pi }}\int_{M}d^{3}x\epsilon ^{\mu \nu \lambda
}\mathrm{Tr}(g^{-1}\partial _{\mu }gg^{-1}\partial _{\nu }gg^{-1}\partial
_{\lambda }g).
\end{equation}%
where the dynamic terms are cancelled and the topological WZW term with time
reversal symmetry is left. To get back the dispersion of the WZW model, a
dynamic term has to be included
\begin{equation}
S_{\mathrm{dyn}}={\frac{1}{8\kappa ^{2}}}\int_{\partial M}dx_{1}dx_{0}%
\mathrm{Tr}(\partial _{1}g\partial _{1}g^{-1}+\partial _{0}g\partial
_{0}g^{-1}).
\end{equation}%
Under the renormalization flow, the fixed point can be reached at ${\frac{1}{%
8\kappa ^{2}}}={\frac{k}{16\pi }}$, leading to the full action of the (1+1)D
$SO(3)$ WZW model (\ref{SO3}).

Moreover, our doubled Chern-Simons model(\ref{DCS}) is essentially a
response action. However, if we regard it as the effective action of a
dynamic gauge field, it should describe two-dimensional topologically
ordered phases with time reversal and parity symmetries, however their
boundary excitations are not necessarily gapless.\cite{Freedman-2003} It has
been shown that each (1+1)D critical field theory of the WZW model
corresponds to a (2+1)D topologically ordered phase, which can be described
by a topological gauge theory or topological spin theory, depending on the
parity of level $k$. It should be mentioned that these results are also
applicable to the $SU(2)$ level-$k$ WZW models, where there is no distinct
difference between even $k$ and odd $k$.

Furthermore, if we regard the symmetry group as one of the chiral symmetry,
namely $SO(3)_{L}$, the resulting effective bulk theory is given by the $%
SO(3)$ Chern-Simons topological theory.\cite{LiuWen} The above
classification of the WZW theory can be used to study the (2+1)D $SO(3)$ SPT
phases, which are classified by the third group cohomology $\mathcal{H}%
^{3}(SO(3),U(1))=Z$, equivalent to the fourth cohomology group $%
H^{4}(BSO(3),Z)$. However, from the inverse transgression map, these SPT
phases can be described by the principal chiral model (\ref{PCSO3}), which
is classified by the third cohomology group $H^{3}(SO(3),Z)$ with \textit{%
even} level $k$. These bulk SPT phases are gapped, while the boundary theory
has gapless excitations as long as the $SO(3)$ symmetry is preserved.
However, for the models with an odd level-$k$, the bulk excitations are
still gapped, but we can have a $Z_{2}$ vortex excitation carrying
half-integer spin according to the $SO(3)_{L}$ symmetry.\cite{LiuWen} This
has also been reflected in the boundary theory, which contains both integer
and half-odd integer excitations. The emergence of fractionalized
excitations indicates that the odd level-$k$ SPT phases represent 2D
topologically ordered states with possible nontrivial statistics\cite%
{LevinWen}, for instance, the chiral spin liquids. Therefore, the
classification of the (1+1)D $SO(3)$ WZW models as the boundary theory of
the two-dimensional principal chiral models leads to the above important
prediction.

\textit{Conclusion}.- We have carefully examined the effective field theory
of the Bethe ansatz integrable Heisenberg antiferromagnetic spin chains. The
quantum critical integer spin chains should be characterized by the $SO(3)$
level-$S$ WZW model and divided into two distinct universality classes,
determined by the parity of the spin. These two classes of WZW models
correspond to 2D doubled $SO(3)$ topological Chern-Simons theory or doubled $%
SO(3)$ topological spin theory, respectively. Furthermore, if we adopt the
chiral symmetry $SO(3)_L$, these two classes of WZW theory describe the
boundary excitations of 2D SPT phases or 2D chiral spin liquid phases,
respectively. Therefore, our present work provides a systematic method to
study 2D topological phases from 1D quantum critical theory. Some other
issues, such as the correspondent chiral spin liquid, are under
investigations.

G. M. Zhang would like to thank Dung-Hai Lee for many stimulating
discussions and for sharing his many insights on topological phases. We
thank Xie Chen, Xiao-Gang Wen, and Yong-Shi Wu for helpful discussions. The
work is supported by the NSF-China and the National Program for Basic
Research of MOST, China.


\begin{thebibliography}{99}
\bibitem{Wen-Niu} X. G. Wen and Q. Niu, Phys. Rev. B \textbf{41}, 9377
(1990).

\bibitem{Tsui-1982} D. C. Tsui, H. L. Stormer, and A. C. Gossard, Phys. Rev.
Lett. \textbf{48}, 1559 (1982).

\bibitem{Laughlin-1983} R. B. Laughlin, Phys. Rev. Lett. \textbf{50}, 1359
(1983).

\bibitem{Kitaev-1997} A. Kitaev, Annals of Physics \textbf{303}, 2 (2003).

\bibitem{chen-gu-wen} X. Chen, Z. C. Gu, and X. G. Wen, Phys. Rev. B \textbf{%
82}, 155138 (2010).

\bibitem{Hasan-Kane} M. Z. Hasan and C. L. Kane, Rev. Mod. Phys. \textbf{82}%
, 3045 (2010).

\bibitem{Qi-Zhang} X. L. Qi and S. C. Zhang, Rev. Mod. Phys. \textbf{83},
1057 (2011).

\bibitem{Haldane-1983} F. D. M. Haldane, Phys. Lett. \textbf{93A}, 464
(1983); Phys. Rev. Lett. \textbf{50}, 1153 (1983).

\bibitem{Gu-Wen-2009} Z. C. Gu and X. G. Wen, Phys. Rev. B \textbf{80},
155131 (2009).

\bibitem{Pollmann-2012} F. Pollmann, E. Berg, A. M. Turner, and M. Oshikawa,
Phys. Rev. B \textbf{85}, 075125 (2012).

\bibitem{Chen-Gu-Wen1D} X. Chen, Z. C. Gu, and X. G. Wen, Phys. Rev. B
\textbf{83}, 035107 (2011); \textit{ibid}, Phys. Rev. B \textbf{84}, 235128
(2011).

\bibitem{chen-gu-liu-wen} X. Chen, Z. C. Gu, Z. X. Liu, and X. G. Wen,
arXiv:1106.4772.

\bibitem{Takhtajan} L. Takhtajan, Phys. Lett. \textbf{87A}, 479 (1982).

\bibitem{Babudjian} J. Babujian, Phys. Lett. \textbf{90A}, 479 (1982); Nucl.
Phys. \textbf{B125} 317 (1983).

\bibitem{Affleck-1986} I. Affleck, Phys. Rev. Lett. \textbf{56}, 746 (1986);
ibid, \textbf{56}, 2763 (1986).

\bibitem{Affleck-Haldane} I. Affleck and F. D. M. Haldane, Phys. Rev. B
\textbf{36}, 5291 (1987).

\bibitem{Affleck-1989} I. Affleck, D. Gepner, H. J. Schulz, and T. Ziman, J.
Phys. A: Math. Gen. \textbf{22}, 511 (1989).

\bibitem{Avdeev-1990} L. V. Avdeed, J. Phys. A: Math. Gen. \textbf{23}, L485
(1990).

\bibitem{Dijkgraaf-Witten} R. Dijkgraaf and E. Witten, Commun. Math. Phys.
\textbf{129}, 393 (1990).

\bibitem{Tsvelik-1990} A. M. Tsvelik, Phys. Rev. B \textbf{42}, 10499 (1990).

\bibitem{Xu-Ludwig} C. Xu and A. W. W. Ludwig, arXiv:1112.5303.

\bibitem{note1} The model (\ref{PCSO3}) also has time reversal symmetry $T$,
namely, it is invariant under $i\to -i$, $\tau\to\tau$, $g\to g^{-1}$.
Notice that the symmetry group $SO(3)_L$ and $SO(3)_R$ exchange their roles
under $T$, also see \cite{LiuWen}.

\bibitem{Gepner-Witten-1986} D. Gepner and E. Witten, Nucl. Phys. \textbf{%
B278}, 493 (1986).

\bibitem{Felder-1988} G. Felder, K. Gawedzki, and A. Kupiainen, Nucl. Phys.
\textbf{B299}, 355 (1988).

\bibitem{GabeGannon} M. R. Gaberdiel and T. Gannon, J. High Energy Phys.
\textbf{4}, 030 (2004).

\bibitem{MooreSeiberg} G. Moore and N. Seiberg, Phys. Lett. B 220, 422
(1989).

\bibitem{LiuWen} Z. X. Liu, X. G. Wen, arXiv: 1205.7024.

\bibitem{note2} Since the $SO(3)_{L}$ symmetry is transformed into $%
SO(3)_{R} $ under time reversal transformation $T$, the corresponding gauge
field $A_{\mu }$ is tranformed into $\bar{A}_{\mu }$. With a gauge
transformation $g\rightarrow h(x)g\bar{h}(x)^{-1}$, the fields $A_{\mu }$
varies as $A_{\mu }\rightarrow hA_{\mu }h^{-1}+h\partial _{\mu }h^{-1}$
while $\bar{A}$ varies as $\bar{A}_{\mu }\rightarrow \bar{h}^{-1}\bar{A}%
_{\mu }\bar{h}+\bar{h}^{-1}\partial _{\mu }\bar{h}$. We can expand $A_{\mu
}=\sum_{a}A_{\mu }^{a}S^{a}$ and $\bar{A}_{\mu }=\sum_{a}\bar{A}_{\mu
}^{a}S^{a}$, $a=x,y,z$. Under $T$, we have $S^{x,z}\rightarrow S^{x,z}$, $%
S^{y}\rightarrow -S^{y}$, this means $A_{\mu }^{x,z}\rightarrow \bar{A}_{\mu
}^{x,z}$ and $A_{\mu }^{y}\rightarrow -\bar{A}_{\mu }^{y}$.

\bibitem{Freedman-2003} M. Freedman, C. Nayak, K. Shtengel, K. Walker and Z.
Wang, Ann. Phys., \textbf{310}, 428 (2004).

\bibitem{LevinWen} M. Levin and X. G. Wen, Phys. Rev. Lett., \textbf{96},
110405 (2006).
\end{thebibliography}
\end{document}